\documentclass[twocolumn,preprintnumbers,amsmath,aps]{revtex4}
\usepackage{graphicx}
\begin{document}
%%%%%%%%%%%%%%%%%%%%%%%%%%%%%%%%%%%%%%%%%%%%%%%%%%%%%%%
\title{Theory for the electronic structure of incommensurate twisted bilayer graphene}
%%%%%%%%%%%%%%%%%%%%%%%%%%%%%%%%%%%%%%%%%%%%%%%%%%%%%%%
\author{D. Ghader$^{1, 2}$}\email{Doried.Ghader@univ-lemans.fr}
\author{D. Szcz{\c e}{\' s}niak$^{1}$}\email{dszczesniak@qf.org.qa}
\author{A. Khater$^{1, 2}$}
%%%%%%%%%%%%%%%%%%%%%%%%%%%%%%%%%%%%%%%%%%%%%%%%%%%%%%%
\affiliation{$^{1}$Qatar Energy and Environment Research Institute, Qatar Foundation, PO Box 5825, Doha, Qatar}
\affiliation{$^{2}$Institute of Molecules and Materials UMR 6283 CNRS, University du Maine, 72000 Le Mans, France}
%%%%%%%%%%%%%%%%%%%%%%%%%%%%%%%%%%%%%%%%%%%%%%%%%%%%%%%
\date{\today} 
\begin{abstract}
%%%%%%%%%%%%%%%%%%%%%%%%%%%%%%%%%%%%%%%%%%%%%%%%%%%%%%%

The experimental control over the twist angle in twisted bilayer graphene has not been reported and its realistic structure is most likely incommensurate. In this paper, we develop a tight-binding virtual crystal approximation theory to study the electronic properties in incommensurate twisted bilayer graphene. The theory yields the electronic band structure and the local density of states for any incommensurate twist angle $\theta$ between the graphene sheets. Angle dependent Van Hove singularities are observed in the numerically calculated local density of states. In accord with observations in scanning tunneling microscopy and spectroscopy, our theoretical calculation indicates that the rotation angle between graphene sheets does not result in a significant reduction in the Fermi velocity in comparison with monolayer graphene. The developed theory is quite general and can be applied to investigate the electronic properties in any incommensurate multilayer heterostructures.

%%%%%%%%%%%%%%%%%%%%%%%%%%%%%%%%%%%%%%%%%%%%%%%%%%%%%%%
\end{abstract}
\maketitle
\noindent{\bf PACS:} 73.22.Pr, 73.20.-r, 61.44.Fw\\
{\bf Keywords:} graphene, interfaces electronic states, incommensurate solids
%%%%%%%%%%%%%%%%%%%%%%%%%%%%%%%%%%%%%%%%%%%%%%%%%%%%%%%

The electronic structures of graphene bilayer systems depend considerably on their stacking structure \cite{zhang, li, bao, yankowitz}. The {\it AB} (Bernal) and {\it AA} stacking configurations, for example, exhibit very different electronic properties despite the subtle distinction between their stacking orders. More recently, the introduction of twisted bilayer graphene (tBLG), consisting of two sheets of graphene rotated relative to each other by an angle $\theta$, has dramatically expanded the allotropes of the graphene bilayer and thin films. In particular, the tBLG demonstrated strong twist-dependent electronic spectra and properties \cite{lopes1, mele1, bistritzer1, yan, brihuega, wang, yin}, rendering the twist angle as a unique degree of freedom to modify the electronic properties of graphene multilayers.

Twisted bilayer graphene is usually grown on the $(000\bar{1})$ face of silicon-carbide (SiC) and its structure is theoretically classified as commensurate or incommensurate, sensitively depending on the rotation angle \cite{mele1}. Commensurate tBLG is periodic and can be viewed as a honeycomb superlattice with  more than four atoms per unit cell and larger lattice vectors. On the contrary, the incommensurate tBLG is non-periodic and characterized by structural disorder that changes with the incommensurate twist angle. Commensuration in tBLG occurs theoretically at a discrete infinite set of twist angles \cite{mele1}, with a vanishing probability that a randomly selected twist angle is commensurate. Experimentally fabricated tBLG are therefore most likely incommensurate, especially since the chemical synthesis of tBLG with a controlled rotation angle has not been reported \cite{wang}.

Despite the intensive theoretical effort \cite{lopes1, mele1, bistritzer1}, \cite{mele2, shallcross, hicks, bistritzer2, bistritzer3, lopes2, landgraf, pal, liang, moon, uchida}, a systematic theory to investigate the electronic structure in realistic incommensurate tBLG is still missing and highly indispensable. First-principles calculations in tBLG, for example, are restricted to commensurate tBLG \cite{liang}, \cite{uchida}. Moreover, current tight-binding theories are restricted to the commensurate structures \cite{lopes1, mele1, bistritzer1}, \cite{mele2, shallcross, hicks, bistritzer2, bistritzer3, lopes2, landgraf}, \cite{liang, moon, uchida} or, at their best, to incommensurate situations near commensuration \cite{pal}.

In this Letter, we develop a non-trivial tight-binding virtual crystal approximation (TB-VCA) theory to study the electronic structure in incommensurate tBLG, representing the first systematic study of the electronic properties in this disordered system. The VCA theoretical approach is used in particular to determine the effective medium for a disordered tBLG system, derived in a direct manner from the quasi-infinite set of structural configurations for carbon atoms at the bilayer interface, for a given incommensurate twist angle. The theory hence yields the TB-VCA Hamiltonian matrix which turns out to be quasi-Hermitian as a consequence of the incommensurate twist angle $\theta$. The hermiticity of the TB-VCA Hamiltonian operator is successfully retrieved bytransposing the inner product, and eventually the Hilbert space associated with the incommensurate tBLG system. The theoretical TB-VCA model yields the electronic structure and the local electronic density of states (LDOS) for any incommensurate twist angle $\theta$.

Our TB-VCA theory underlines the pseudo-Hermitian representation of quantum mechanics \cite{bender1, bender2, mostafazadeh1, mostafazadeh2}. Pseudo-Hermitian quantum mechanics is a recent and unconventional approach to quantum mechanics based on the use of non-Hermitian Hamiltonians, whose hermiticity can be retrieved by transposing the inner product defining the system Hilbert space. The pseudo-Hermitian representation of quantum mechanics and the techniques developed in the course of its investigation have found applications in a variety of subjects \cite{mostafazadeh1}, including condensed matter physics, scattering theory, relativistic quantum mechanics, and quantum cosmology.

The TB-VCA theoretical method developed in this work for incommensurate graphene tBLG bilayer structures is quite general, and can be generalized in a straightforward manner to study the electronic structure and optoelectronic properties for any multiple incommensurate twisted graphene heterostructure.

The Letter is organized as follows. We present first the mathematical formalism developed for the TB-VCA theory. The numerical applications are then presented and discussed, and the last part   is dedicated to conclusions and perspectives.

\begin{figure}[ht!]
\centering
\includegraphics[width=\columnwidth]{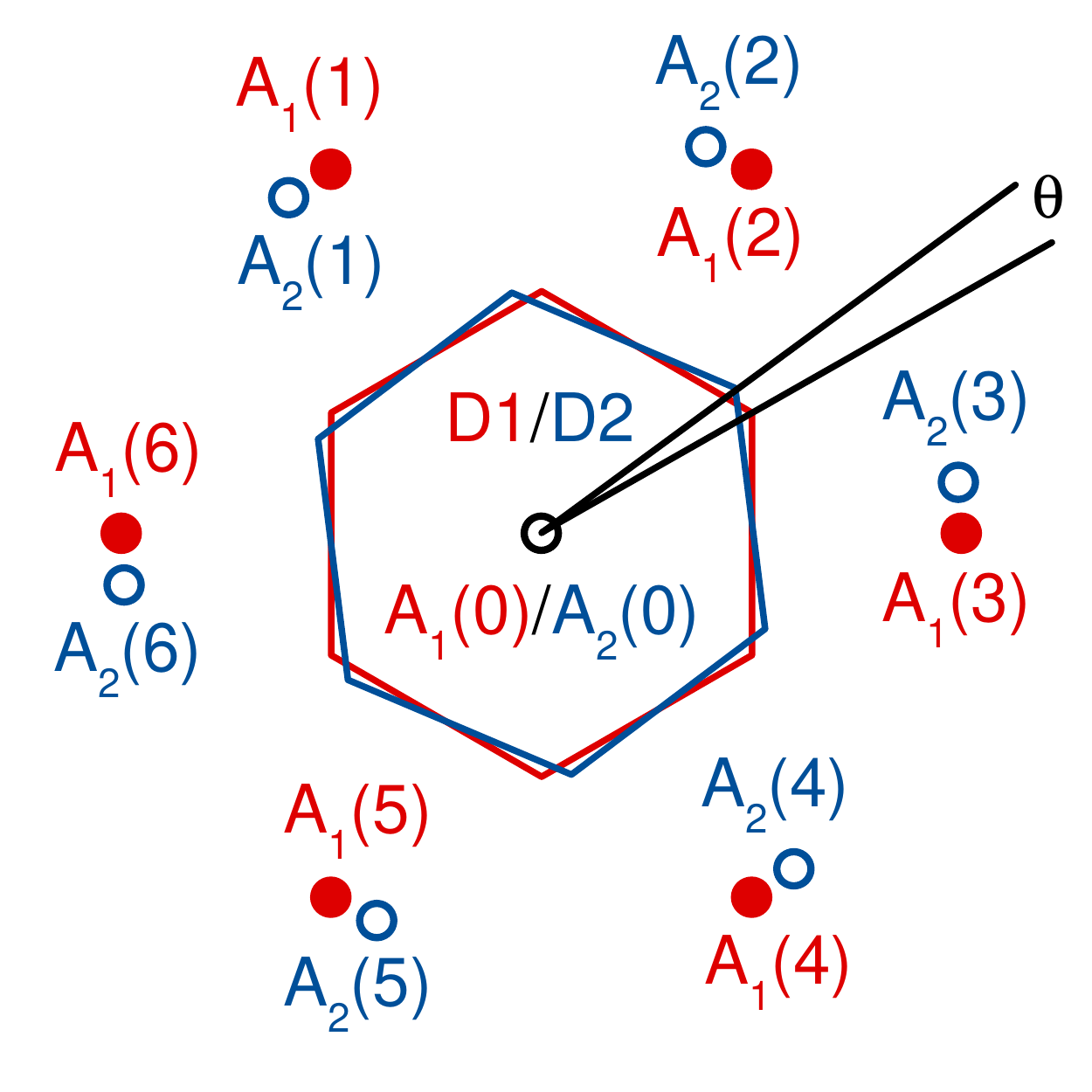}
\caption{The figure shows the numerically determined region $D_{1}$ in layer 1 (region $D_{2}$ in layer 2) which serve to simulate the quasi-infinite ensembles of the structural configurations and interlayer bonds, at the incommensurate interface of the bilayer tBLG system, for different twist angles $\theta$. This is done with respect to the referential sites $A_{1}(0)$ ($A_{2}(0)$) for the A sublattice. Similar figures may be schematically prepared for the other three configurations for the interactions of A and B sublattices in the two layers of the tBLG system.}
\label{fig1}
\end{figure}

The Tight-Binding (TB) representation is presented next for the particular tBLG problem. Consider the tBLG as composed of two graphene layers, denoted layer 1 and 2, where layer 2 is rotated by an angle $\theta$ with respect to layer 1. See Fig.\ref{fig1} for the schematic representation. Monolayer graphene is known to be composed of two sublattices named $A$ and $B$. We hence refer to the sublattices in tBLG as $A_{i}$ and $B_{i}$ sublattices, with $i=1$ and 2 for layers 1 and 2 respectively. Atoms in $A_{i}$ and $B_{i}$ sublattices are respectively labeled $A_{i}(n)$ and $B_{i}(n)$ with $n=0, 1, 2, ...$. In the absence of the periodic structure in incommensurate tBLG, our TB-VCA model considers referential coupled sites in each layer, denoted by $\{A_{1}(0), B_{1}(0)\}$ and $\{A_{2}(0), B_{2}(0)\}$, in layers 1 and 2 respectively.

In the VCA treatment of the disordered tBLG, the effective medium is determined as a direct system average and the reciprocal-space techniques can hence be applied for the VCA periodic effective medium. Considering a single $p_{z}$-orbital per site, the total electronic wave function may hence be written as
\begin{eqnarray}
\label{eq1}
\Psi_{\bf k}({\bf r})&=&C_{A_{1}} \Psi_{{\bf k}; A_{1}} ({\bf r}) + C_{B_{1}} \Psi_{{\bf k}; B_{1}} ({\bf r}) \nonumber \\
&+& C_{A_{2}} \Psi_{{\bf k}; A_{2}} ({\bf r}) + C_{B_{2}} \Psi_{{\bf k}; B_{2}} ({\bf r}),
\end{eqnarray}
\noindent where $\Psi_{{\bf k}; \alpha}({\bf r})=[1/\sqrt{N}]\sum_{{\bf r}_\alpha}e^{i {\bf k}.{\bf r}_{\alpha}}\Phi_{\alpha}({\bf r}-{\bf r}_{\alpha})$, $\alpha=A_{1}, B_{1}, A_{2}, B_{2}$, is the TB-VCA Bloch wave function for the $\alpha$ sublattice. Bold letters are used to denote vector quantities. $N$ is the number of unit cells in the crystal, $C_{\alpha}$ is the $\alpha$-sublattice contribution to the total wave function, $\Phi_{\alpha}$ denotes the $2p_{z}$ atomic orbital, ${\bf r}_{\alpha}$ is the position of an $\alpha$ type atom with respect to a chosen origin, and ${\bf k}$ is the wave vector.

We denote by $V$ the vector space spanned by the basis $\{\Psi_{{\bf k}; A_{1}},\Psi_{{\bf k}; B_{1}},\Psi_{{\bf k}; A_{2}},\Psi_{{\bf k}; B_{2}}\}$. The standard inner product $\left<.|.\right>$ defined on $V$ is given by $\left< \chi | \phi \right> = \iiint \chi^{*}({\bf r}) \phi({\bf r}) {\rm d} {\bf r}$ for any $\chi$ and $\phi$ in $V$. The vector space $V$ endowed by the inner product $\left<.|.\right>$ defines a Hilbert space $\mathcal{H}$. This may be used to describe the incommensurate tBLG via the TB-VCA Hamiltonian operator $\hat{H}$ satisfying the Schr{\" o}dinger equation
\begin{equation}
\label{eq2}
\hat{H}\Psi_{\bf k} ({\bf r}) = E({\bf k}) \Psi_{\bf k} ({\bf r}),
\end{equation}
where $E({\bf k})$ denotes the energy eigenvalue. It is important to note that the Hilbert space used to describe the incommensurate tBLG is not unique and may be changed by transposing the corresponding inner product.

As in a standard tight-binding model, the energy eigenvalue can be determined by solving the secular equation
\begin{equation}
\label{eq3}
{\rm det} \left[ H - E^{\lambda} ({\bf k}) S \right] = 0.
\end{equation}
The label $\lambda$ denotes the energy bands. $H$ and $S$ denote respectively the Hamiltonian and the overlap matrices with matrix entities $S_{\alpha \beta}= \left< \Psi_{{\bf k}; \alpha} | \Psi_{{\bf k}; \beta} \right> = \iiint {\rm d}^{3} {\bf r} \Psi_{{\bf k}; \alpha}^{*} ({\bf r}) \Psi_{{\bf k}; \beta} ({\bf r})$ and $H_{\alpha \beta}= \left< \Psi_{{\bf k}; \alpha} | \hat{H} | \Psi_{{\bf k}; \beta} \right> = \iiint {\rm d}^{3} {\bf r} \Psi_{{\bf k}; \alpha}^{*} ({\bf r}) \hat{H} \Psi_{{\bf k}; \beta}({\bf r})$; $\alpha, \beta \in \{ A_{1}, B_{1}, A_{2}, B_{2} \}$. We note that the definition of the matrices $H$ and $S$ follow directly the definition of the inner product in the Hilbert space.

It is convenient to write each of the $H$ and $S$ in terms of 4 block matrices as follows
\begin{equation}
\label{eq4}
H=\left[ \begin{array}{c c}
H_{1 1} (2 \times 2) & H_{1 2} (2 \times 2) \\
H_{2 1} (2 \times 2) & H_{2 2} (2 \times 2)
\end{array} \right],
\end{equation}
\noindent where
\begin{equation}
\label{eq5}
H_{m n}(2 \times 2)=\left[ \begin{array}{c c}
H_{A_{m} A_{n}} & H_{A_{m} B_{n}} \\
H_{B_{m} A_{n}} & H_{B_{m} B_{n}}
\end{array} \right],
\end{equation}
and similarly for the overlap matrix $S$.

The restriction to the single $p_{z}$ - orbital per site yields $pp\pi$ in-plane and $pp\sigma$ interlayer bonds in the incommensurate tBLG system. In our calculations, we include up to the third nearest neighbors in-plane hopping and overlap integrals, and the explicit form for the in-plane block matrices $H_{1 1}$, $H_{2 2}$, $S_{1 1}$ and $S_{2 2}$ can be found in reference  \cite{kundu}. We further neglect the interlayer overlap integrals, justified by the large perpendicular distance $d_{\perp}$ between the two graphene layers $\left(d_{\perp} \approx 3.35 {\rm \AA} \right)$. Consequently $S_{1 2} = S_{2 1}=0$. Note in this case that  $S_{1 1}$ and $S_{2 2}$ are invariant under the twist angle.  We are hence left with the determination of the interlayer block matrices $H_{1 2}$ and $H_{2 1}$.

Through a numerical based statistical approach, we then analyzed the disordered interface morphology and the interlayer interactions in incommensurate tBLG. In particular, we have carefully investigated the interlayer interactions between each of the referential sites  $\{ A_{i}(0), B_{i}(0) \}$ from one layer interacting with sites  $\{ A_{j}(n), B_{j}(n) \}$ in the other layer ($j \neq i$).

Our study demonstrates the existence of a quasi-infinite set of configurations for carbon atoms at the tBLG bilayer interface, and hence a quasi-continuum set of interlayer bonds. This quasi-continuum set of interlayer bonds requires a theoretical approach which yields the interlayer hopping integral as a function of the bond lengths. In our work, we choose the environment dependent tight-binding model, developed initially by Tang {\it et al.} \cite{tang}, and optimized by Landgraf {\it et al.} \cite{landgraf}, for the case of the tBLG.

The key idea in our TB-VCA approach is hence to determine an ensemble average for the interlayer hopping integrals, and another average over the quasi-infinite ensemble of structural configurations. In this context, it is very useful to explore the possible statistical identities over the  structural disorder of the  incommensurate tBLG system. For example, our numerical analysis for the interlayer interactions demonstrates identical ensembles of interlayer bonds and configurations, whether we chose the referential sites  $A_{1}(0)$ or $B_{1}(0)$ on layer 1 to interact with sites  $\{ A_{2}(n), B_{2}(n) \}$ in layer 2 or vice-versa. This analysis over the amorphous-like structural disorder of the incommensurate tBLG system, yields the following important identities
\begin{equation}
\label{eq6}
\bar{H}_{A_{1} A_{2}}=\bar{H}_{A_{1} B_{2}}=\bar{H}_{B_{1} A_{2}}=\bar{H}_{B_{1} B_{2}},
\end{equation}
and
\begin{equation}
\label{eq7}
\bar{H}_{A_{2} A_{1}}=\bar{H}_{A_{2} B_{1}}=\bar{H}_{B_{2} A_{1}}=\bar{H}_{B_{2} B_{1}}.
\end{equation}
Note that the matrix  $\bar{H}_{A_{1}A_{2}}$ ($\bar{H}_{A_{2}A_{1}}$) entities correspond respectively to the interlayer interactions between the referential sites $A_{1}(0)$ [$A_{2}(0)$] with the $A_{2}$ [$A_{1}$] sublattices. This is also generically the case for the other identities of Eq.\ref{eq6} and Eq.\ref{eq7}. 

The TB-VCA Hamiltonian matrix may hence be written from Eq. \ref{eq4}, Eq. \ref{eq6}, and Eq. \ref{eq7}, in a reduced form as
\begin{equation}
\label{eq8}
\bar{H}=\left[ \begin{array}{c c}
\bar{H}_{1 1} & \bar{H}_{A_{1} A_{2}}Q \\
\bar{H}_{A_{2} A_{1}}Q & \bar{H}_{2 2}
\end{array} \right],
\end{equation}
with
\begin{equation}
\label{eq9}
Q=\left[ \begin{array}{c c}
1 & 1 \\
1 & 1
\end{array} \right],
\end{equation}

In what follows the TB virtual crystal approximation (TB-VCA) theory will be given. In order to proceed and determine the required TB-VCA ensemble averages, and consequently the matrix entities $\bar{H}_{A_{1}A_{2}}$ and $\bar{H}_{A_{2}A_{1}}$, it is essential at this stage to simulate the ensembles of interlayer bonds and structural configurations. In Fig.\ref{fig1}, we present the numerically determined regions $D_{1}$ (in layer 1) and $D_{2}$ (in layers 2) which serve this purpose. For example, by varying the position of atom $A_{2}(0)$ such that its projection spans region $D_{1}$, we formally simulate the ensemble of interlayer bonds between the referential site $A_{1}(0)$ and sublattice $A_{2}$ for the ensemble of the structural configurations of the incommensurate interface. With this simulation, and using the environment dependent tight binding model approach, atom $A_{1}(0)$ is found to interact with atoms $\{ A_{2} (n), n=0, ..., 6 \}$ of the $A_{2}$ sublattices. 

The matrix entity $\bar{H}_{A_{1} A_{2}}$ may hence be determined in the TB-VCA approach as
\begin{eqnarray}
\label{eq10}
\bar{H}_{A_{1} A_{2}}&=&\sum_{n=0}^{6} \bar{t} \left[ A_{1}(0) A_{2}(n) \right] \nonumber \\
&\times&{\rm exp} \left[ i ({\bf r}_{A_{2}(n)} - {\bf r}_{A_{2}(0)}.{\bf k} \right],
\end{eqnarray}
with

\begin{widetext}
\begin{equation}
\label{eq11}
\bar{t} \left[A_{i}(0) A_{j}(0) \right]=\frac{\iint_{D_{1}} t \left(||{\bf r}_{A_{j}(n)}-{\bf r}_{A_{i}(0)}|| \right) \Gamma \left(R_{\rm max} - ||{\bf r}_{A_{j}(n)}-{\bf r}_{A_{i}(0)}|| \right) {\rm d}x {\rm d}y}{\iint_{D_{1}}\Gamma \left(R_{\rm max} - ||{\bf r}_{A_{j}(n)}-{\bf r}_{A_{i}(0)}|| \right) {\rm d}x {\rm d}y}
\end{equation}
\end{widetext}

\begin{figure*}[ht]
\centering
\includegraphics[width=\textwidth]{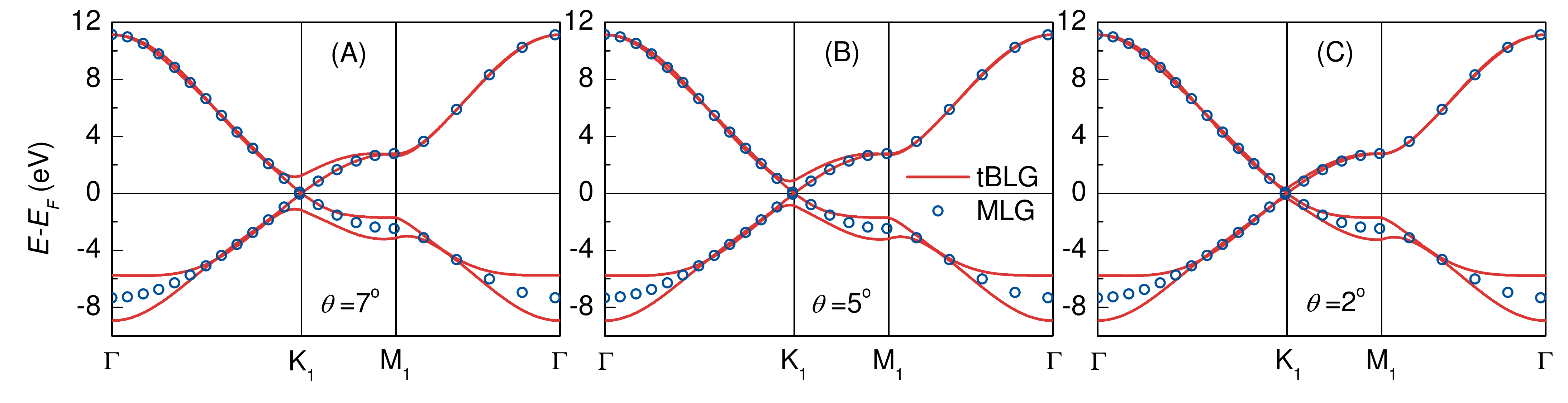}
\caption{The numerically calculated TB-VCA electronic structures for the tBLG system with incommensurate twist angles from left to right: $\theta = 7^{\circ}$ (A), $5^{\circ}$ (B) and $2^{\circ}$ (C), plotted along the high symmetry axes $\{\Gamma K_{1}, K_{1} M_{1}, M_{1} \Gamma \}$ in the first BZ of layer 1. Saddle points are observed in these electronic structures owing to the overlap between the Dirac cones. The circles represent the electronic band structure of a single graphene monolayer. The Fermi velocity near $K_{1}$ for the tBLG system is observed to be identical to that of mass-less Dirac fermions in monolayer graphene.}
\label{fig2}
\end{figure*}

In Eq.\ref{eq11}, $||\hspace{0.1cm}||$ designates the norm of a given vector, ${\bf r}_{\alpha}$ is the position of atom $\alpha$ with respect to an arbitrary origin, $\Gamma$ denotes the Heaviside step function. The $R_{\rm max}$ is the maximum interlayer separation beyond which the interlayer interaction in neglected, $t$ is the distance-dependent interlayer hopping integral determined by the environment dependent tight binding model. $\bar{t} \left[A_{1}(0) A_{2}(n) \right]$ is the average interlayer hopping integral between atoms $A_{1} (0)$ and $A_{2}(n)$.

Due to symmetry (see Fig.\ref{fig1}), the averaged hopping integrals $\bar{t} \left[A_{1}(0) A_{2}(n) \right]$ for $n=1, ..., 6$ are all identical. This yields hence
\begin{eqnarray}
\label{eq12}
\bar{H}_{A_{1} A_{2}}&=& \bar{t}\left[A_{1}(0)A_{2}(0) \right] + 2 \bar{t}\left[A_{1}(0)A_{2}(1)\right] \nonumber \\
&\times& \sum_{n=1}^{3} {\rm cos} \left[\left({\bf r}_{A_{2}(n)}-{\bf r}_{A_{1}(0)}\right).{\bf k} \right].
\end{eqnarray}
The same approach yields the matrix entity $\bar{H}_{A_{2}A_{1}}$ as follows
\begin{eqnarray}
\label{eq13}
\bar{H}_{A_{2} A_{1}}&=& \bar{t}\left[A_{2}(0)A_{1}(0) \right] + 2 \bar{t}\left[A_{2}(0)A_{1}(1)\right] \nonumber \\
&\times& \sum_{n=1}^{3} {\rm cos} \left[\left({\bf r}_{A_{1}(n)}-{\bf r}_{A_{2}(0)}\right).{\bf k} \right].
\end{eqnarray}
Note that $\bar{t}\left[A_{1}(0)A_{2}(0) \right]$ is invariant with twist angle, whereas $\bar{t}\left[A_{1}(0)A_{2}(n) \right]$ varies as an ensemble average with the twist angle. Also due to the bilayer inversion symmetry, $\bar{t}\left[A_{1}(0)A_{2}(m) \right]$ = $\bar{t}\left[A_{2}(0)A_{1}(m) \right]$ for $m$ = 0, and $m$ = 1, 2,.., 6.

By this, we complete the derivation of the Hamiltonian matrix $\bar{H}$, which turns out to be a quasi-hermitian matrix  \cite{mostafazadeh1}, \cite{mostafazadeh2}. The hermiticity of the Hamiltonian operator $\hat{H}$ describing the incommensurate tBLG in the TB-VCA approach may be retrieved by transposing the inner product defined on the function vector space V.

This is done by introducing a new inner product $\left<.|.\right>_{\eta}$ on $V$, determined by a metric operator $\hat{\eta}$ defined by $\hat{\eta}\Psi_{{\bf k}; \alpha_{1}}= \left( 1+ \delta \right)\Psi_{{\bf k}; \alpha_{1}}$  and $\hat{\eta}\Psi_{{\bf k}; \alpha_{2}}= \left( 1+ \delta^{-1} \right)\Psi_{{\bf k}; \alpha_{2}}$ with $\delta=\bar{H}_{A_{2} A_{1}}/\bar{H}_{A_{1} A_{2}}$ and $\alpha=A$ or $B$. This yields a Hermitian Hamiltonian matrix $\bar{H}'$ given as
\begin{equation}
\label{eq14}
\bar{H}'=\left[ \begin{array}{c c}
\left(1 + \delta\right)\bar{H}_{1 1} & \left(\bar{H}_{A_{1} A_{2}} + \bar{H}_{A_{2} A_{1}} \right) Q \\
\left(\bar{H}_{A_{1} A_{2}} + \bar{H}_{A_{2} A_{1}} \right)Q & \left(1 + 1/\delta\right) \bar{H}_{2 2}
\end{array} \right].
\end{equation}
The quasi-Hermitian and Hermitian representations of the TB-VCA Hamiltonian operator $\hat{H}$ are formally equivalent \cite{mostafazadeh1}, \cite{mostafazadeh2}, and yield the same results. The electronic structure for the incommensurate tBLG may hence be determined in a formal manner by solving the secular equation \ref{eq3}. 

The TB-VCA numerical applications are presented next for different tBLG systems. The numerical values of the onsite energy, the in-plane hopping integrals, and the overlap  integrals are adapted from reference \cite{kundu}. The numerically calculated averaged interlayer hopping integrals are presented in Table 1. The numerical value of $R_{\rm max}$ used in this calculation is $R_{\rm max} \approx 3.8 {\rm \AA}$.

\begin{table}[ht]
\caption{\label{tab1} The ensemble averaged interlayer hopping integrals calculated numerically.}
\begin{ruledtabular}
\begin{tabular}{c c c}

 & $\bar{t} \left[A_{1}(0) A_{2}(0) \right]$ (eV) & $\bar{t} \left[A_{1}(0) A_{2}(1) \right]$ (eV)\\

\hline

$\theta = 7^{\circ}$ & 0.2907 & 0.1348\\

$\theta = 5^{\circ}$ & 0.2907 & 0.1343\\

$\theta = 2^{\circ}$ & 0.2907 & 0.1340\\

$\theta = 1^{\circ}$ & 0.2907 & 0.1339 \\

\end{tabular}
\end{ruledtabular}
\end{table}

Our numerical calculations demonstrate the existence of two overlapped Dirac cones centered at the Dirac points $K_{1}$ and $K_{2}$ in the Brillouin zones of layers 1 and 2 respectively. In Figs.$\ref{fig2}$, we plot the electronic structure of incommensurate tBLG (solid red curves) along the high symmetry axes $\{\Gamma K_{1}, K_{1} M_{1}, M_{1} \Gamma \}$ in the first BZ of layer 1. The figures from left to right correspond to the twist angles $\theta = 7^{\circ}$, $5^{\circ}$ and $2^{\circ}$.  The open circles in the figures correspond to the electronic structure of graphene monolayer. The figures show that the Fermi velocity below the Van Hove singularities (VHS), is found in our theory to be identical to that of mass-less Dirac fermions in monolayer graphene. This result is in accordance with recent scanning tunneling microscopy and spectroscopy studies on incommensurate tBLG \cite{yan}, \cite{yin}.

\begin{figure}[ht]
\centering
\includegraphics[width=\columnwidth]{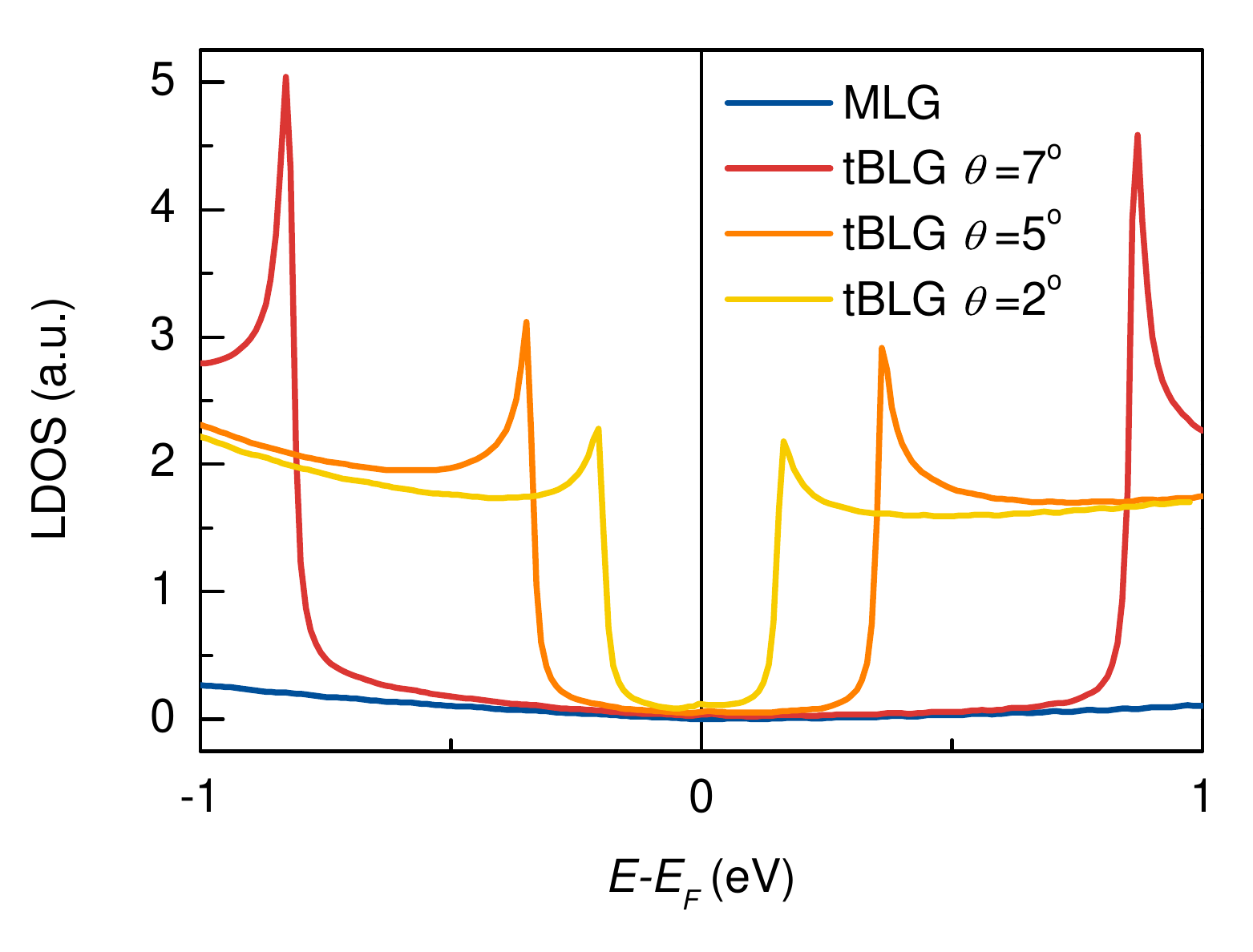}
\caption{The curves represent the numerically calculated LDOS in the TB-VCA model, in the neighborhood  of the Dirac point $K_{1}$ for the incommensurate twist angles $\theta = 7^{\circ}$, $5^{\circ}$ and $2^{\circ}$, where the wider is the separation between the LDOS peaks the bigger is the twist angle. The tBLG LDOS present Van Hove singularities due to the overlap between the Dirac cones at $K_{1}$ and $K_{2}$. The low lying, almost flat, curve corresponds to the LDOS of single graphene monolayer. }
\label{fig3}
\end{figure}

The electronic structures in Figs.$\ref{fig2}$ further demonstrate the existence of saddle points, induced from the band-overlap between the two Dirac cones. These saddle points generate the VHS peaks in the local density of states (LDOS) which have been observed experimentally \cite{li}, \cite{bistritzer1, yan, brihuega}, \cite{yin}. The numerically calculated TB-VCA LDOS along the high symmetry axes $\{\Gamma K_{1}, K_{1} M_{1}, M_{1} \Gamma \}$ are presented in Figs.$\ref{fig3}$ for the twist angles $\theta = 5^{\circ}$, $2^{\circ}$ and $1^{\circ}$. In accordance with experimental results, the saddle points and consequently the VHSs are observed to shift to lower energies when the twist angle is reduced. 

In conclusion we present the first systematic study of the electronic properties in incommensurate twisted bilayer graphene. In particular, we have developed and applied the TB-VCA theory to determine the electronic band structure and local density of states in tBLG for any incommensurate twist angle $\theta$. The theory establishes for the disordered incommensurate tBLG system an averaged periodic VCA effective medium, deduced directly from the quasi-continuum set of interlayer bounds and the quasi-infinite set of structural configurations characterizing the disordered system. The theory yields the quasi-Hermitian and Hermitian equivalent representations for the TB-VCA Hamiltonian operator describing the incommensurate tBLG. The numerical calculation determines consequently the electronic band structure and the angle dependent VHS in the LDOS for any incommensurate twist angle. The Fermi velocity is found to be identical to that of mass-less Dirac fermions in monolayer graphene.

The current TB-VCA theory is quite general and can be applied to study the electronic structure in incommensurate twisted multilayer graphene and  layered heterostructures. Concerning perspectives, the authors are currently developing a dynamic variant of the non-local coherent potential approximation which fully describes the decay of electronic excitations in the tBLG system, adopting the current TB-VCA as a basic reference.

%%%%%%%%%%%%%%%%%%%%%%%%%%%%%%%%%%%%%%%%%%%%%%%%%%%%%%%
\bibliographystyle{apsrev}
\bibliography{manuscript}

\begin{thebibliography}{28}
\expandafter\ifx\csname natexlab\endcsname\relax\def\natexlab#1{#1}\fi
\expandafter\ifx\csname bibnamefont\endcsname\relax
  \def\bibnamefont#1{#1}\fi
\expandafter\ifx\csname bibfnamefont\endcsname\relax
  \def\bibfnamefont#1{#1}\fi
\expandafter\ifx\csname citenamefont\endcsname\relax
  \def\citenamefont#1{#1}\fi
\expandafter\ifx\csname url\endcsname\relax
  \def\url#1{\texttt{#1}}\fi
\expandafter\ifx\csname urlprefix\endcsname\relax\def\urlprefix{URL }\fi
\providecommand{\bibinfo}[2]{#2}
\providecommand{\eprint}[2][]{\url{#2}}

\bibitem[{\citenamefont{Zhang et~al.}(2009)\citenamefont{Zhang, Tang, Girit,
  Hao, Martin, Zettl, Crommie, Shen, and Wang}}]{zhang}
\bibinfo{author}{\bibfnamefont{Y.}~\bibnamefont{Zhang}},
  \bibinfo{author}{\bibfnamefont{T.~T.} \bibnamefont{Tang}},
  \bibinfo{author}{\bibfnamefont{C.}~\bibnamefont{Girit}},
  \bibinfo{author}{\bibfnamefont{Z.}~\bibnamefont{Hao}},
  \bibinfo{author}{\bibfnamefont{M.~C.} \bibnamefont{Martin}},
  \bibinfo{author}{\bibfnamefont{A.}~\bibnamefont{Zettl}},
  \bibinfo{author}{\bibfnamefont{M.~F.} \bibnamefont{Crommie}},
  \bibinfo{author}{\bibfnamefont{Y.~R.} \bibnamefont{Shen}}, \bibnamefont{and}
  \bibinfo{author}{\bibfnamefont{F.}~\bibnamefont{Wang}},
  \bibinfo{journal}{Nature} \textbf{\bibinfo{volume}{459}},
  \bibinfo{pages}{820} (\bibinfo{year}{2009}).

\bibitem[{\citenamefont{Li et~al.}(2010)\citenamefont{Li, Luican, {Lopes dos
  Santos}, {Castro Neto}, Reina, Kong, and Andrei}}]{li}
\bibinfo{author}{\bibfnamefont{G.}~\bibnamefont{Li}},
  \bibinfo{author}{\bibfnamefont{A.}~\bibnamefont{Luican}},
  \bibinfo{author}{\bibfnamefont{J.~M.~B.} \bibnamefont{{Lopes dos Santos}}},
  \bibinfo{author}{\bibfnamefont{A.~H.} \bibnamefont{{Castro Neto}}},
  \bibinfo{author}{\bibfnamefont{A.}~\bibnamefont{Reina}},
  \bibinfo{author}{\bibfnamefont{J.}~\bibnamefont{Kong}}, \bibnamefont{and}
  \bibinfo{author}{\bibfnamefont{E.~Y.} \bibnamefont{Andrei}},
  \bibinfo{journal}{Nat. Phys.} \textbf{\bibinfo{volume}{6}},
  \bibinfo{pages}{109} (\bibinfo{year}{2010}).

\bibitem[{\citenamefont{Bao et~al.}(2011)\citenamefont{Bao, Jing, {Velasco Jr},
  Lee, Liu, Tran, Standley, Aykol, Cronin, Smirnov et~al.}}]{bao}
\bibinfo{author}{\bibfnamefont{W.}~\bibnamefont{Bao}},
  \bibinfo{author}{\bibfnamefont{L.}~\bibnamefont{Jing}},
  \bibinfo{author}{\bibfnamefont{J.}~\bibnamefont{{Velasco Jr}}},
  \bibinfo{author}{\bibfnamefont{Y.}~\bibnamefont{Lee}},
  \bibinfo{author}{\bibfnamefont{G.}~\bibnamefont{Liu}},
  \bibinfo{author}{\bibfnamefont{D.}~\bibnamefont{Tran}},
  \bibinfo{author}{\bibfnamefont{B.}~\bibnamefont{Standley}},
  \bibinfo{author}{\bibfnamefont{M.}~\bibnamefont{Aykol}},
  \bibinfo{author}{\bibfnamefont{S.~B.} \bibnamefont{Cronin}},
  \bibinfo{author}{\bibfnamefont{D.}~\bibnamefont{Smirnov}},
  \bibnamefont{et~al.}, \bibinfo{journal}{Nat. Phys.}
  \textbf{\bibinfo{volume}{7}}, \bibinfo{pages}{948} (\bibinfo{year}{2011}).

\bibitem[{\citenamefont{Yankowitz et~al.}(2013)\citenamefont{Yankowitz, Wang,
  Lau, and LeRoy}}]{yankowitz}
\bibinfo{author}{\bibfnamefont{M.}~\bibnamefont{Yankowitz}},
  \bibinfo{author}{\bibfnamefont{F.}~\bibnamefont{Wang}},
  \bibinfo{author}{\bibfnamefont{C.~N.} \bibnamefont{Lau}}, \bibnamefont{and}
  \bibinfo{author}{\bibfnamefont{B.~J.} \bibnamefont{LeRoy}},
  \bibinfo{journal}{Phys. Rev. B} \textbf{\bibinfo{volume}{87}},
  \bibinfo{pages}{165102} (\bibinfo{year}{2013}).

\bibitem[{\citenamefont{{Lopes dos Santos} et~al.}(2007)\citenamefont{{Lopes
  dos Santos}, Peres, and {Castro Neto}}}]{lopes1}
\bibinfo{author}{\bibfnamefont{J.~M.~B.} \bibnamefont{{Lopes dos Santos}}},
  \bibinfo{author}{\bibfnamefont{N.~M.~R.} \bibnamefont{Peres}},
  \bibnamefont{and} \bibinfo{author}{\bibfnamefont{A.~H.} \bibnamefont{{Castro
  Neto}}}, \bibinfo{journal}{Phys. Rev. B} \textbf{\bibinfo{volume}{99}},
  \bibinfo{pages}{256802} (\bibinfo{year}{2007}).

\bibitem[{\citenamefont{Mele}(2010)}]{mele1}
\bibinfo{author}{\bibfnamefont{E.~J.} \bibnamefont{Mele}},
  \bibinfo{journal}{Phys. Rev. B} \textbf{\bibinfo{volume}{81}},
  \bibinfo{pages}{161405(R)} (\bibinfo{year}{2010}).

\bibitem[{\citenamefont{R. and MacDonald}(2011{\natexlab{a}})}]{bistritzer1}
\bibinfo{author}{\bibfnamefont{B.}~\bibnamefont{R.}} \bibnamefont{and}
  \bibinfo{author}{\bibfnamefont{A.~H.} \bibnamefont{MacDonald}},
  \bibinfo{journal}{PNAS} \textbf{\bibinfo{volume}{108}},
  \bibinfo{pages}{12233} (\bibinfo{year}{2011}{\natexlab{a}}).

\bibitem[{\citenamefont{Yan et~al.}(2012)\citenamefont{Yan, Liu, Dou, Meng,
  Feng, Chu, Zhang, Liu, Nie, and He}}]{yan}
\bibinfo{author}{\bibfnamefont{W.}~\bibnamefont{Yan}},
  \bibinfo{author}{\bibfnamefont{M.}~\bibnamefont{Liu}},
  \bibinfo{author}{\bibfnamefont{R.~F.} \bibnamefont{Dou}},
  \bibinfo{author}{\bibfnamefont{L.}~\bibnamefont{Meng}},
  \bibinfo{author}{\bibfnamefont{L.}~\bibnamefont{Feng}},
  \bibinfo{author}{\bibfnamefont{Z.~D.} \bibnamefont{Chu}},
  \bibinfo{author}{\bibfnamefont{Y.}~\bibnamefont{Zhang}},
  \bibinfo{author}{\bibfnamefont{Z.}~\bibnamefont{Liu}},
  \bibinfo{author}{\bibfnamefont{J.~C.} \bibnamefont{Nie}}, \bibnamefont{and}
  \bibinfo{author}{\bibfnamefont{L.}~\bibnamefont{He}}, \bibinfo{journal}{Phys.
  Rev. Lett} \textbf{\bibinfo{volume}{109}}, \bibinfo{pages}{126801}
  (\bibinfo{year}{2012}).

\bibitem[{\citenamefont{Brihuega et~al.}(2012)\citenamefont{Brihuega, Mallet,
  Gonz{\' a}lez-Herrero, {Trambly de Laissardi{\` e}re}, Ugeda, Magaud, G{\'
  o}mez-Rodr{\' i}guez, Yndur{\' a}in, and Veuillen}}]{brihuega}
\bibinfo{author}{\bibfnamefont{I.}~\bibnamefont{Brihuega}},
  \bibinfo{author}{\bibfnamefont{P.}~\bibnamefont{Mallet}},
  \bibinfo{author}{\bibfnamefont{H.}~\bibnamefont{Gonz{\' a}lez-Herrero}},
  \bibinfo{author}{\bibfnamefont{G.}~\bibnamefont{{Trambly de Laissardi{\`
  e}re}}}, \bibinfo{author}{\bibfnamefont{M.~M.} \bibnamefont{Ugeda}},
  \bibinfo{author}{\bibfnamefont{L.}~\bibnamefont{Magaud}},
  \bibinfo{author}{\bibfnamefont{J.~M.} \bibnamefont{G{\' o}mez-Rodr{\'
  i}guez}}, \bibinfo{author}{\bibfnamefont{F.}~\bibnamefont{Yndur{\' a}in}},
  \bibnamefont{and} \bibinfo{author}{\bibfnamefont{J.~Y.}
  \bibnamefont{Veuillen}}, \bibinfo{journal}{Phys. Rev. Lett}
  \textbf{\bibinfo{volume}{109}}, \bibinfo{pages}{196802}
  (\bibinfo{year}{2012}).

\bibitem[{\citenamefont{Wang et~al.}(2014)\citenamefont{Wang, Su, Wu, Nie, Lu,
  Wang, McCarty, Pei, Robles-Hernandez, Hadjiev et~al.}}]{wang}
\bibinfo{author}{\bibfnamefont{Y.}~\bibnamefont{Wang}},
  \bibinfo{author}{\bibfnamefont{Z.}~\bibnamefont{Su}},
  \bibinfo{author}{\bibfnamefont{W.}~\bibnamefont{Wu}},
  \bibinfo{author}{\bibfnamefont{S.}~\bibnamefont{Nie}},
  \bibinfo{author}{\bibfnamefont{X.}~\bibnamefont{Lu}},
  \bibinfo{author}{\bibfnamefont{H.}~\bibnamefont{Wang}},
  \bibinfo{author}{\bibfnamefont{K.}~\bibnamefont{McCarty}},
  \bibinfo{author}{\bibfnamefont{S.~S.} \bibnamefont{Pei}},
  \bibinfo{author}{\bibfnamefont{F.}~\bibnamefont{Robles-Hernandez}},
  \bibinfo{author}{\bibfnamefont{V.~G.} \bibnamefont{Hadjiev}},
  \bibnamefont{et~al.}, \bibinfo{journal}{Nanotechnology}
  \textbf{\bibinfo{volume}{25}}, \bibinfo{pages}{335201}
  (\bibinfo{year}{2014}).

\bibitem[{\citenamefont{Yin et~al.}(2015)\citenamefont{Yin, Qiao, Xu, Dou, Nie,
  and He}}]{yin}
\bibinfo{author}{\bibfnamefont{L.~J.} \bibnamefont{Yin}},
  \bibinfo{author}{\bibfnamefont{J.~B.} \bibnamefont{Qiao}},
  \bibinfo{author}{\bibfnamefont{R.}~\bibnamefont{Xu}},
  \bibinfo{author}{\bibfnamefont{R.~F.} \bibnamefont{Dou}},
  \bibinfo{author}{\bibfnamefont{J.~C.} \bibnamefont{Nie}}, \bibnamefont{and}
  \bibinfo{author}{\bibfnamefont{L.}~\bibnamefont{He}},
  \bibinfo{journal}{arXiv:1410.1621v3 [cond-mat.mes-hall]}
  (\bibinfo{year}{2015}).

\bibitem[{\citenamefont{Mele}(2011)}]{mele2}
\bibinfo{author}{\bibfnamefont{E.~J.} \bibnamefont{Mele}},
  \bibinfo{journal}{Phys. Rev. B} \textbf{\bibinfo{volume}{84}},
  \bibinfo{pages}{235439} (\bibinfo{year}{2011}).

\bibitem[{\citenamefont{Shallcross et~al.}(2010)\citenamefont{Shallcross,
  Sharma, Kandelaki, and Pankratov}}]{shallcross}
\bibinfo{author}{\bibfnamefont{S.}~\bibnamefont{Shallcross}},
  \bibinfo{author}{\bibfnamefont{S.}~\bibnamefont{Sharma}},
  \bibinfo{author}{\bibfnamefont{E.}~\bibnamefont{Kandelaki}},
  \bibnamefont{and} \bibinfo{author}{\bibfnamefont{O.~A.}
  \bibnamefont{Pankratov}}, \bibinfo{journal}{Phys. Rev. B}
  \textbf{\bibinfo{volume}{81}}, \bibinfo{pages}{165105}
  (\bibinfo{year}{2010}).

\bibitem[{\citenamefont{Hicks et~al.}(2011)\citenamefont{Hicks, Sprinkle,
  Shepperd, Wang, Tejeda, Taleb-Ibrahimi, Bertran, {Le F{\` e}vre}, {de Heer},
  Berger et~al.}}]{hicks}
\bibinfo{author}{\bibfnamefont{J.}~\bibnamefont{Hicks}},
  \bibinfo{author}{\bibfnamefont{M.}~\bibnamefont{Sprinkle}},
  \bibinfo{author}{\bibfnamefont{K.}~\bibnamefont{Shepperd}},
  \bibinfo{author}{\bibfnamefont{F.}~\bibnamefont{Wang}},
  \bibinfo{author}{\bibfnamefont{A.}~\bibnamefont{Tejeda}},
  \bibinfo{author}{\bibfnamefont{A.}~\bibnamefont{Taleb-Ibrahimi}},
  \bibinfo{author}{\bibfnamefont{F.}~\bibnamefont{Bertran}},
  \bibinfo{author}{\bibfnamefont{P.}~\bibnamefont{{Le F{\` e}vre}}},
  \bibinfo{author}{\bibfnamefont{W.~A.} \bibnamefont{{de Heer}}},
  \bibinfo{author}{\bibfnamefont{C.}~\bibnamefont{Berger}},
  \bibnamefont{et~al.}, \bibinfo{journal}{Phys. Rev. B}
  \textbf{\bibinfo{volume}{83}}, \bibinfo{pages}{205403}
  (\bibinfo{year}{2011}).

\bibitem[{\citenamefont{R. and MacDonald}(2010)}]{bistritzer2}
\bibinfo{author}{\bibfnamefont{B.}~\bibnamefont{R.}} \bibnamefont{and}
  \bibinfo{author}{\bibfnamefont{A.~H.} \bibnamefont{MacDonald}},
  \bibinfo{journal}{Phys. Rev. B} \textbf{\bibinfo{volume}{81}},
  \bibinfo{pages}{245412} (\bibinfo{year}{2010}).

\bibitem[{\citenamefont{R. and MacDonald}(2011{\natexlab{b}})}]{bistritzer3}
\bibinfo{author}{\bibfnamefont{B.}~\bibnamefont{R.}} \bibnamefont{and}
  \bibinfo{author}{\bibfnamefont{A.~H.} \bibnamefont{MacDonald}},
  \bibinfo{journal}{Phys. Rev. B} \textbf{\bibinfo{volume}{84}},
  \bibinfo{pages}{035440} (\bibinfo{year}{2011}{\natexlab{b}}).

\bibitem[{\citenamefont{{Lopes dos Santos} et~al.}(2012)\citenamefont{{Lopes
  dos Santos}, Peres, and {Castro Neto}}}]{lopes2}
\bibinfo{author}{\bibfnamefont{J.~M.~B.} \bibnamefont{{Lopes dos Santos}}},
  \bibinfo{author}{\bibfnamefont{N.~M.~R.} \bibnamefont{Peres}},
  \bibnamefont{and} \bibinfo{author}{\bibfnamefont{A.~H.} \bibnamefont{{Castro
  Neto}}}, \bibinfo{journal}{Phys. Rev. B} \textbf{\bibinfo{volume}{86}},
  \bibinfo{pages}{155449} (\bibinfo{year}{2012}).

\bibitem[{\citenamefont{Landgraf et~al.}(2013)\citenamefont{Landgraf,
  Shallcross, T{\" u}rschmann, Weckbecker, and Pankratov}}]{landgraf}
\bibinfo{author}{\bibfnamefont{W.}~\bibnamefont{Landgraf}},
  \bibinfo{author}{\bibfnamefont{S.}~\bibnamefont{Shallcross}},
  \bibinfo{author}{\bibfnamefont{K.}~\bibnamefont{T{\" u}rschmann}},
  \bibinfo{author}{\bibfnamefont{D.}~\bibnamefont{Weckbecker}},
  \bibnamefont{and}
  \bibinfo{author}{\bibfnamefont{O.}~\bibnamefont{Pankratov}},
  \bibinfo{journal}{Phys. Rev. B} \textbf{\bibinfo{volume}{87}},
  \bibinfo{pages}{075433} (\bibinfo{year}{2013}).

\bibitem[{\citenamefont{Pal et~al.}(2014)\citenamefont{Pal, Carter, and
  Kindermann}}]{pal}
\bibinfo{author}{\bibfnamefont{H.~K.} \bibnamefont{Pal}},
  \bibinfo{author}{\bibfnamefont{S.}~\bibnamefont{Carter}}, \bibnamefont{and}
  \bibinfo{author}{\bibfnamefont{M.}~\bibnamefont{Kindermann}},
  \bibinfo{journal}{arXiv:1409.1971v1 [cond-mat.mes-hall]}
  (\bibinfo{year}{2014}).

\bibitem[{\citenamefont{Liang}(2014)}]{liang}
\bibinfo{author}{\bibfnamefont{Y.}~\bibnamefont{Liang}}, Ph.D. thesis,
  \bibinfo{school}{Washington University in St. Louis} (\bibinfo{year}{2014}).

\bibitem[{\citenamefont{Moon et~al.}(2014)\citenamefont{Moon, Son, and
  Koshino}}]{moon}
\bibinfo{author}{\bibfnamefont{P.}~\bibnamefont{Moon}},
  \bibinfo{author}{\bibfnamefont{Y.~W.} \bibnamefont{Son}}, \bibnamefont{and}
  \bibinfo{author}{\bibfnamefont{M.}~\bibnamefont{Koshino}},
  \bibinfo{journal}{Phys. Rev. B} \textbf{\bibinfo{volume}{90}},
  \bibinfo{pages}{155427} (\bibinfo{year}{2014}).

\bibitem[{\citenamefont{Uchida et~al.}(2014)\citenamefont{Uchida, Furuya,
  Iwata, and Oshiyama}}]{uchida}
\bibinfo{author}{\bibfnamefont{K.}~\bibnamefont{Uchida}},
  \bibinfo{author}{\bibfnamefont{S.}~\bibnamefont{Furuya}},
  \bibinfo{author}{\bibfnamefont{J.~I.} \bibnamefont{Iwata}}, \bibnamefont{and}
  \bibinfo{author}{\bibfnamefont{A.}~\bibnamefont{Oshiyama}},
  \bibinfo{journal}{Phys. Rev. B} \textbf{\bibinfo{volume}{90}},
  \bibinfo{pages}{155451} (\bibinfo{year}{2014}).

\bibitem[{\citenamefont{Bender and Boettcher}(1998)}]{bender1}
\bibinfo{author}{\bibfnamefont{C.~M.} \bibnamefont{Bender}} \bibnamefont{and}
  \bibinfo{author}{\bibfnamefont{S.}~\bibnamefont{Boettcher}},
  \bibinfo{journal}{Phys. Rev. Lett.} \textbf{\bibinfo{volume}{80}},
  \bibinfo{pages}{5243} (\bibinfo{year}{1998}).

\bibitem[{\citenamefont{Bender}(2007)}]{bender2}
\bibinfo{author}{\bibfnamefont{C.~M.} \bibnamefont{Bender}},
  \bibinfo{journal}{Rep. Prog. Phys.} \textbf{\bibinfo{volume}{70}},
  \bibinfo{pages}{947} (\bibinfo{year}{2007}).

\bibitem[{\citenamefont{Mostafazadeh}(2010)}]{mostafazadeh1}
\bibinfo{author}{\bibfnamefont{A.}~\bibnamefont{Mostafazadeh}},
  \bibinfo{journal}{Int. J. Geom. Methods Mod. Phys.}
  \textbf{\bibinfo{volume}{7}}, \bibinfo{pages}{1191} (\bibinfo{year}{2010}).

\bibitem[{\citenamefont{Mostafazadeh}(2009)}]{mostafazadeh2}
\bibinfo{author}{\bibfnamefont{A.}~\bibnamefont{Mostafazadeh}},
  \bibinfo{journal}{Pramana} \textbf{\bibinfo{volume}{73}},
  \bibinfo{pages}{269} (\bibinfo{year}{2009}).

\bibitem[{\citenamefont{Kundu}(2011)}]{kundu}
\bibinfo{author}{\bibfnamefont{R.}~\bibnamefont{Kundu}}, \bibinfo{journal}{Mod.
  Phys. Lett. B} \textbf{\bibinfo{volume}{25}}, \bibinfo{pages}{163}
  (\bibinfo{year}{2011}).

\bibitem[{\citenamefont{Tang et~al.}(1996)\citenamefont{Tang, Wang, Chan, and
  Ho}}]{tang}
\bibinfo{author}{\bibfnamefont{M.~S.} \bibnamefont{Tang}},
  \bibinfo{author}{\bibfnamefont{C.~Z.} \bibnamefont{Wang}},
  \bibinfo{author}{\bibfnamefont{C.~T.} \bibnamefont{Chan}}, \bibnamefont{and}
  \bibinfo{author}{\bibfnamefont{K.~M.} \bibnamefont{Ho}},
  \bibinfo{journal}{Phys. Rev. B} \textbf{\bibinfo{volume}{53}},
  \bibinfo{pages}{979} (\bibinfo{year}{1996}).

\end{thebibliography}
%%%%%%%%%%%%%%%%%%%%%%%%%%%%%%%%%%%%%%%%%%%%%%%%%%%%%%%
\end{document}